\documentclass[12pt,draftclsnofoot,onecolumn]{IEEEtran}
\IEEEoverridecommandlockouts
\usepackage{booktabs}
\usepackage[T1]{fontenc}
\makeatletter
\newsavebox{\@tabnotebox}

\makeatother	
\usepackage{graphicx}
\usepackage{floatrow}

\newfloatcommand{figurebox}{figure}[\nocapbeside][\dimexpr(\textwidth-\columnsep)/2\relax]
\newfloatcommand{tablebox}{table}[\nocapbeside][\dimexpr(\textwidth-\columnsep)/2\relax]

\makeatother

\newif\ifhavebib

\usepackage{blindtext}
\usepackage{graphicx}
\usepackage{graphicx}
\usepackage{array}
\usepackage{url}
\usepackage{algorithmic}
\usepackage[cmex10]{amsmath}
\usepackage{ifpdf}
\usepackage{amssymb,amsmath}
\usepackage{empheq}
\usepackage{stfloats}   
\usepackage{textcomp}
\usepackage{cite}
\usepackage{multirow}
\usepackage{diagbox}
\usepackage{subcaption}
\let\oldFootnote\footnote
\newcommand\nextToken\relax

\renewcommand\footnote[1]{%
    \oldFootnote{#1}\futurelet\nextToken\isFootnote}

\newcommand\isFootnote{%
    \ifx\footnote\nextToken\textsuperscript{,}\fi}

\usepackage[ruled,vlined,lined,ruled,boxed]{algorithm2e}
\usepackage[dvipsnames,usenames]{color}
\usepackage[normalem]{ulem}
\usepackage{amsthm}

\usepackage{graphicx}
\usepackage{ifpdf}
\usepackage{cite}
\usepackage{enumerate}
\usepackage{enumitem}

\usepackage{array}
\usepackage{mdwmath}
\usepackage{mdwtab}
\usepackage{eqparbox}
\usepackage{bm}

\usepackage{setspace}
\usepackage{booktabs}
\usepackage[subnum]{cases}
\usepackage{rotating}

\usepackage{listings} 
\definecolor{Red}{rgb}{1,0,0}
\definecolor{Blue}{rgb}{0,0,1}
\definecolor{Olive}{rgb}{0.41,0.55,0.13}
\definecolor{Green}{rgb}{0,1,0}
\definecolor{MGreen}{rgb}{0,0.8,0}
\definecolor{DGreen}{rgb}{0,0.55,0}
\definecolor{Yellow}{rgb}{1,1,0}
\definecolor{Cyan}{rgb}{0,1,1}
\definecolor{Magenta}{rgb}{1,0,1}
\definecolor{Orange}{rgb}{1,.5,0}
\definecolor{Violet}{rgb}{.5,0,.5}
\definecolor{Purple}{rgb}{.75,0,.25}
\definecolor{Brown}{rgb}{.75,.5,.25}
\definecolor{Grey}{rgb}{.5,.5,.5}

\newcommand{\boxhead}[5]{
   \pagestyle{myheadings}
   \thispagestyle{plain}
   \setcounter{page}{1}
   \noindent
   \begin{center}
   \framebox{
      \vbox{\vspace{2mm}
    \hbox to 6.28in { {\bf #1 \hfill} }
       \vspace{6mm}
       \hbox to 6.28in { {\Large \hfill \bf #2  \hfill} }
       \vspace{6mm}
       \hbox to 6.28in { {\it #3 #4 \hfill  #5} }
      \vspace{2mm}}
   }
   \end{center}
   \markboth{#5 -- #2}{#5 -- #2}
   \vspace*{4mm}
}

\usepackage[colorlinks=true, urlcolor=black,  linkcolor=black, citecolor=black]{hyperref}
\theoremstyle{definition}

\theoremstyle{remark}

\theoremstyle{definition}

\DeclarePairedDelimiterX{\infdivx}[2]{(}{)}{%
	#1\;\delimsize\|\;#2%
}

\DeclarePairedDelimiter{\norm}{\lVert}{\rVert}
\DeclarePairedDelimiter{\abs}{\lvert}{\rvert}

\DeclareMathOperator*{\argmin}{\mathop{\arg\min}}
\def\tr{\mathop{\rm tr}\nolimits}%
\def\diag{\mathop{\rm diag}\nolimits}%
\def\rank{\mathop{\rm rank}\nolimits}%
\newcommand{\Vv}{{\bf V}}
\newcommand{\Rv}{{\bf R}}

\newcommand{\rv}{{\bf r}}

\newcommand{\Gv}{{\bf G}}

\newcommand{\Iv}{{\bf I}}
\newcommand{\fv}{{\bf f}}
\newcommand{\gv}{{\bf g}}

\newcommand{\uv}{{\bf u}}

\newcommand{\hv}{{\bf h}}
\newcommand{\sv}{{\bf s}}

\newcommand{\nv}{{\bf n}}

\newcommand{\wv}{{\bf w}}

\newcommand{\thetav}{\boldsymbol \theta}

\def\e{\epsilon}

\DeclareMathOperator\E{E}

 \def\E{\mathbb{E}}

\def\de \mathrm{d}

\newcommand{\CN}{\mathcal{CN}}

\newcommand\eg{e.g.,\xspace}
\newcommand\ie{i.e.,\xspace}
\def\textiid{i.i.d.\@\xspace}

\newcommand\iid{\ifmmode\text{ i.i.d. } \else \textiid \fi}

\newcommand{\Complex}{\mathbb{C}}
\newcommand{\Real}{\mathbb{R}}

\newcommand{\beqs}{\begin{equation*}}
\newcommand{\eeqs}{\end{equation*}}
\newcommand{\beq}{\begin{equation}}
\newcommand{\eeq}{\end{equation}}
\begin{document}



\setitemize{listparindent=\parindent,partopsep=0pt,topsep=-0.25ex}
\setenumerate{fullwidth,itemindent=\parindent,listparindent=\parindent,itemsep=0ex,partopsep=0pt,parsep=0ex}

\havebibtrue
\title{
	CSIT-Free Model Aggregation for Federated Edge Learning via Reconfigurable Intelligent Surface
}
\author{
	Hang~Liu, Xiaojun~Yuan,~\IEEEmembership{Senior Member,~IEEE,}
	and~Ying-Jun~Angela~Zhang,~\IEEEmembership{Fellow,~IEEE}
	\thanks{H. Liu and Y.-J. A. Zhang are with the Department of Information Engineering, The Chinese University of Hong Kong, Shatin, New Territories, Hong Kong (e-mail: lh117@ie.cuhk.edu.hk; yjzhang@ie.cuhk.edu.hk). 
		
		X. Yuan is with the Center for Intelligent Networking and Communications, the University of Electronic Science and Technology of China, Chengdu, China (e-mail: xjyuan@uestc.edu.cn).}
}
\maketitle

\begin{abstract}
	We study over-the-air model aggregation in federated edge learning (FEEL) systems, where channel state information at the transmitters (CSIT) is assumed to be unavailable. 
	We leverage the \emph{reconfigurable intelligent surface} (RIS) technology to align the cascaded channel coefficients for \emph{CSIT-free} model aggregation. To this end, we jointly optimize the RIS and the receiver by minimizing the aggregation error under the channel alignment constraint. We then develop a difference-of-convex algorithm for the resulting non-convex optimization. Numerical experiments on image classification show that the proposed method is able to achieve a similar learning accuracy as the state-of-the-art CSIT-based solution, demonstrating the efficiency of our approach in combating the lack of CSIT.

\end{abstract}
\section{Introduction}
With the explosive increase in the number of connected devices at mobile edge networks, machine learning (ML) over a vast volume of data at edge devices has attracted considerable research attention. Federated edge learning (FEEL)  \cite{FEDSGD} has been proposed to enable distributed model training at the network edge. In FEEL, edge devices simultaneously train local models by exploiting local data and periodically upload these models to a parameter server (PS, \eg a base station) to compute a global model (a.k.a. model aggregation). This global model is then sent back to the devices to perform  training in the next round. 

The communication between edge devices and the PS, particularly in model aggregation, is the main bottleneck of FEEL \cite{FEDSGD}. This is because simultaneous model uploading from a large number of devices through unreliable wireless channels incurs large time delay and high bandwidth costs. 
Much research in recent years has been focused on communication protocol design for FEEL model aggregation.  Notably, over-the-air model aggregation has been proposed for concurrent model uploading through a shared channel \cite{GZhu_BroadbandAircomp}. By multiplying scaling factors to the transmitted signals, the wireless fading can be combated, and the local models can be coherently aligned at the PS. 
Over-the-air model aggregation is so far achieved via proper transmission scaling at edge devices, which critically relies on the availability of channel state information (CSI) at the transmitter side (CSIT).
In practice, CSI is acquired at the PS and fed back to the devices through downlink control channels \cite{MIMO_DTse}. As a result, the inevitable error in CSIT feedback brings additional distortions in signal alignment. Moreover, frequent updates of CSIT are needed when channel states change, which leads to high delay and thus slows down the FEEL process. To address these challenges, the authors in \cite{amiri2020blind} proposed a CSIT-free model aggregation solution by exploiting the massive multiple-input multiple-output (MIMO) technique. It shows that, as the number of receive antennas tends to infinity, the inter-user interference in model aggregation diminishes even when devices transmit signals without transmission scaling. 

Although massive MIMO has the potential to achieve CSIT-free FEEL, it requires the deployment of large antenna arrays and may lead to  high power consumption. The \emph{reconfigurable intelligent surface} (RIS) technology has emerged as a green substitute of massive MIMO \cite{yuan2020reconfigurableintelligentsurface}. Specifically, a RIS is a thin sheet comprising a large number of low-cost elements that can induce independent and \emph{passive} phase shifts on the incident signals without the need for radio-frequency chains. {The power consumption of a RIS element is much lower than that of an active antenna \cite{wu2020intelligent}, and hence integrating RISs to existing MIMO systems can improve the energy efficiency \cite{LIS_CHuang}. }
Furthermore, recent studies in \cite{wang2020federated,liu2020reconfigurable} show that RISs can efficiently reduce the model aggregation error and accelerate the convergence of FEEL with the availability of CSIT. 

Inspired by the above developments, we leverage the RIS for FEEL model aggregation with neither CSIT nor a large receive antenna array. Specifically, we consider a single-RIS-assisted FEEL system with a single-antenna PS.
We assume perfect CSI at the PS and no CSIT at the edge devices. Unlike the existing RIS-assisted FEEL algorithms \cite{wang2020federated,liu2020reconfigurable} that rely on CSIT to align signals, we propose to use the RIS phase shifts to adjust the channel coefficients for over-the-air model aggregation. To this end, we constrain the cascaded channel coefficients, as functions of the RIS phase shifts, to be proportional to the corresponding weights of the local models. We derive an expression of the associated model aggregation error under this constraint and formulate the system design problem as minimizing the model aggregation error. We develop a difference-of-convex (DC) algorithm to solve the resulting non-convex problem. 
Numerical results show that the proposed algorithm achieves substantial accuracy improvements compared with the method without RISs. Furthermore, it is shown that, with a sufficiently large RIS, the proposed CSIT-free algorithm performs as well as the CSIT-based one in terms of learning accuracy.

\section{System Model}
\subsection{FEEL Framework}

We consider an FEEL framework that minimizes a loss function $F(\wv;\mathcal{D})$, where $\wv\in\Real^{d\times 1}$ is the $d$-dimensional model parameter vector to be optimized, and $\mathcal{D}$ is the set of available training data. Suppose that $\mathcal{D}$ is distributed over $M$ edge devices. Denote the local dataset at the $m$-th device by $\mathcal{D}_m$. We assume $\bigcup_{m=1}^M\mathcal{D}_m=\mathcal{D}$ and $\mathcal{D}_m\bigcap\mathcal{D}_{m^\prime}=\emptyset,\forall m,m^\prime$. Following \cite{FEDSGD} we have
\begin{align}\label{e01}
	F(\wv;\mathcal{D})&=\sum_{m=1}^M p_m F_m(\wv;\mathcal{D}_m),
\end{align}
where $F_m(\wv;\mathcal{D}_m)\triangleq \frac{1}{|\mathcal{D}_m|}\sum_{\uv\in \mathcal{D}_m}f(\wv;\uv)$; $|\mathcal{D}_m|$ is the cardinality of $\mathcal{D}_m$;
$p_m\triangleq |\mathcal{D}_m|/|\mathcal{D}|$; and $f(\wv;\uv)$ is the local loss function with respect to (w.r.t.) training sample $\uv$.

Here, we adopt the federated averaging algorithm (FedAvg) in \cite{FEDSGD} to minimize \eqref{e01}. At round $t=1,\cdots,T$, the current global model $\wv_t$ is first broadcast to all the edge devices.
Then, every device updates a local model $\wv_{m,t}$ based on $\wv_t$ by mini-batch stochastic gradient descent. Specifically, each local dataset $\mathcal{D}_m$ is portioned into $B$ mini-batches. For each epoch from $1$ to $E$, we sequentially update $\wv_{m,t}\leftarrow \wv_{m,t}-\eta \nabla F_m(\wv_{m,t};\mathcal{B})$ for every mini-batch $\mathcal{B}$, where $\eta$ is the learning rate and $ \nabla F_m(\wv_{m,t};\mathcal{B})$ is the gradient of $F_m(\cdot)$ at $\wv=\wv_{m,t}$. 
Finally, each device computes the local model change $\Delta\wv_{m,t}\triangleq \wv_{m,t}-\wv_{t}$ and uploads it to the PS {through wireless channels. The PS intends to estimate the global model change $\sum_m p_m \Delta\wv_{m,t}$, and compute $\wv_{t+1}\leftarrow \wv_t+\widehat{ \sum_{m=1}^M p_m\Delta\wv_{m,t}}$, where $\widehat{ \sum_{m=1}^M p_m\Delta\wv_{m,t}}$ is the estimate of ${ \sum_{m=1}^M p_m\Delta\wv_{m,t}}$.}
The model uploading and model aggregation in updating $\wv_{t+1}$ act as the main bottleneck that limits the learning performance \cite{GZhu_BroadbandAircomp}. 
The estimation error in model aggregation jeopardizes the learning convergence \cite{liu2020reconfigurable}. 
In the subsequent subsection, we adopt the over-the-air model aggregation technique and use the RIS to enhance the communication efficiency. 

\subsection{RIS-Assisted Communication for Model Aggregation}
\begin{figure}[!t]
	\centering
	\includegraphics[width=2.6in]{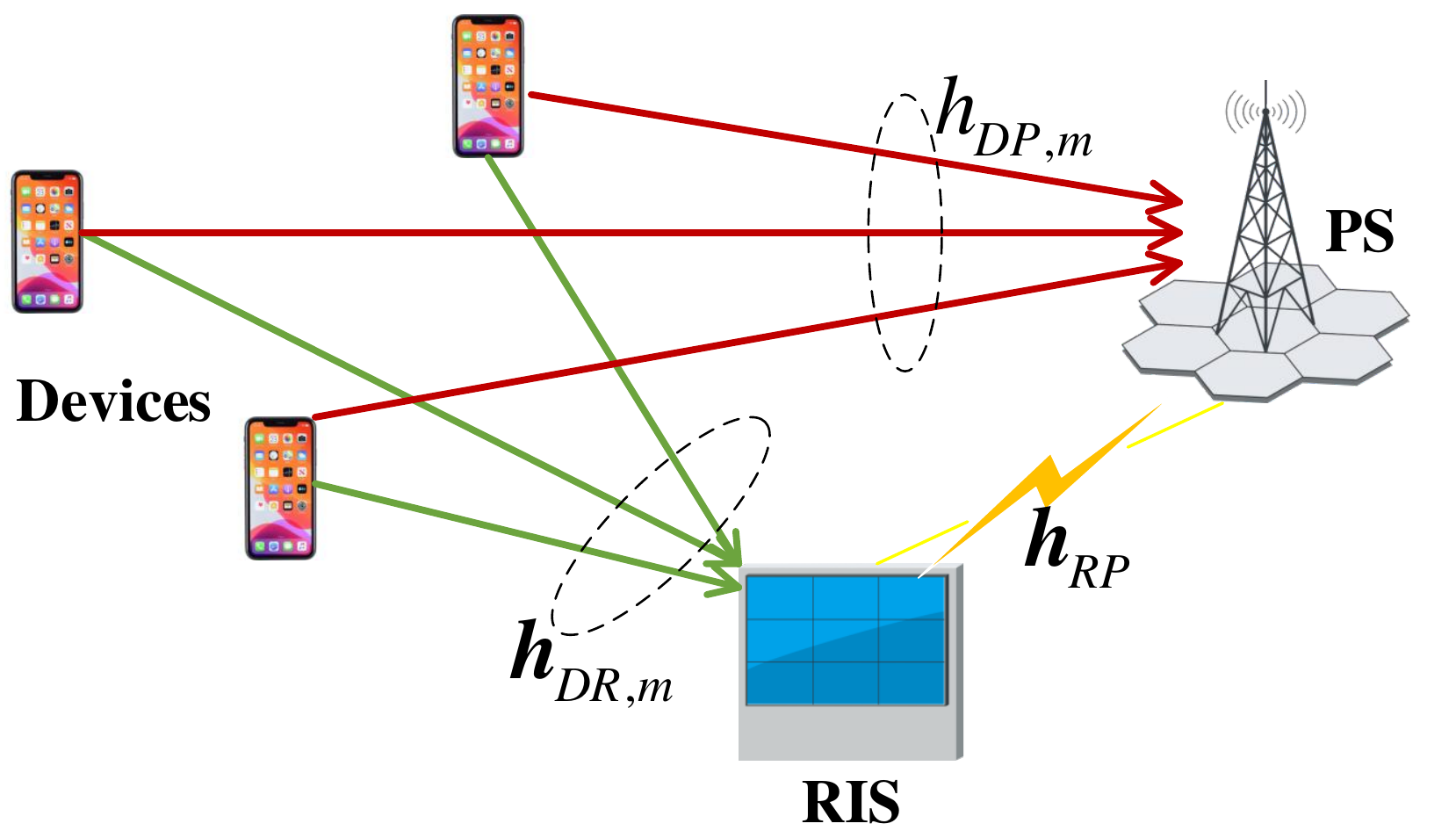}
	\caption{The RIS-assisted communication system.}
	\label{figsystem}
\end{figure}

Consider the single-cell communication system depicted in  Fig. \ref{figsystem}, where a base station acts as the PS to serve $M$ edge devices for FEEL. We assume that the PS and the devices all have one antenna. A RIS with $L$ elements is deployed to assist the communication between the PS and the devices, where each RIS element induces an independent phase shift on the incident signals. We keep the RIS phase shifts invariant during the FEEL training process and denote the phase-shift vector as $\thetav\in \Complex^{L\times 1}$ with $\abs{\theta_l}=1$ for $l=1,2,\cdots,L$.
{For simplicity of exposition}, we assume that the channel coefficients remain invariant during the training process by following \cite{GZhu_BroadbandAircomp,FL_1,liu2020reconfigurable}.
Let $h_{DP,m}\in \Complex$, $\hv_{RP}\in \Complex^{L\times 1}$, and $\hv_{DR,m}\in \Complex^{L\times 1}$, $m=1,2,\cdots,M$, denote the direct $m$-th-device-PS, the RIS-PS, and the $m$-th-device-RIS channel coefficients, respectively. 
The \emph{effective} $m$-th-device-PS channel $h_m(\thetav)$ is the superposition of the direct channel and the RIS cascaded channel as 
\begin{align}
	h_m(\thetav)\triangleq h_{DP,m}+\hv_{RP}^T\diag(\thetav)\hv_{DR,m}=h_{DP,m}+\gv_m^T\thetav,
\end{align}
where $(\cdot)^T$ is the transpose; $\diag(\thetav)$ is the diagonal matrix with the diagonal entries specified by $\thetav$; and $\gv_m^T\triangleq \hv_{RP}^T\diag(\hv_{DR,m})\in \Complex^{1\times L}$. 

In over-the-air model aggregation \cite{GZhu_BroadbandAircomp}, the devices upload $\{\Delta\wv_{m,t}\}$ over the same time-frequency resource, and the PS estimates the weighted sum $\widehat{\sum_m p_m \Delta\wv_{m,t}}$ by exploiting the channel superposition property. The details are described as follows. For brevity, we omit the round index $t$ in the sequel.  

First, the devices encode the local update vectors $\{\gv_m\in \Real^d\}$ into normalized symbol vectors $\{\sv_m\in\Complex^{d\times 1}\}$ to ensure that $\E[\sv_m\sv_m^H]=\Iv_d,\forall m$, and $\E[\sv_m\sv_{m^\prime}^H]={\bf 0}_d,\forall m,m^\prime$, where $(\cdot)^H$ is the conjugate transpose; and $\Iv_d$ (or ${\bf 0}_d$) is the $d\times d$ identity (or zero) matrix.
The normalization operation can be found, \eg in \cite{GZhu_BroadbandAircomp}. Then, at any transmission time slot $i=1,2,\cdots,d$ of a training round, the devices send their signals $\{s_m[i],\forall m\}$ to the PS simultaneously. The corresponding received signal at the PS, denoted by $y[i]$, is given by
\begin{align}
	y[i]&=\sum_{m=1}^M h_m(\thetav)b_m s_m[i]+n[i],
\end{align}   
where $b_m\in\Complex$ is the complex-valued transmit scalar at the $m$-th device, and  $n[i]\sim\CN(0,\sigma^2)$ is the additive white Gaussian noise (AWGN) following a zero-mean Gaussian distribution with variance $\sigma^2$.
We consider an individual  transmit power constraint as
\begin{align}\label{eq03}
	\E[|b_ms_m[i]|^2]=|b_m|^2\leq P_0, \forall m,i,
\end{align}   
where $P_0$ is the maximum transmit power at each device.

At the PS, we directly estimate the desired weighted sum $\sum_m p_m s_m[i]$ by a de-noising receive scalar  $c\in\Complex$. Specifically, the estimate of $\sum_m p_m s_m[i]$, denoted by $\widehat r[i]$, is given by
\begin{align}\label{eq05}
	\widehat r[i]=\frac{y[i]}{c}=\sum_{m=1}^M \frac{h_m(\thetav)b_m} {c}s_m[i]+
	\frac{n[i]}{c}.
\end{align}
{After collecting $\hat \rv=[\hat r[1],\cdots,\hat r[d]]$, our target $\widehat{ \sum_{m=1}^M p_m\Delta\wv_{m,t}}$ is obtained by applying the de-normalization process in \cite{GZhu_BroadbandAircomp} to $\hat \rv$.}
The estimation performance can be tracked by the mean-square error (MSE) between $\sum_m p_m s_m[i]$ and $\widehat r[i]$ as
\begin{align}\label{eq06}
	\text{MSE}=\E\left[\Big|\sum_m p_m s_m[i]-\widehat r[i]\Big|^2\right],
\end{align}
where the expectation is taken w.r.t. $n[i]$.

\section{Model Aggregation Without CSIT}
{As shown in \cite{liu2020reconfigurable}, FEEL is guaranteed to converge with a sufficiently small model aggregation MSE in \eqref{eq06}.}
To ensure the convergence of FEEL, we need to optimize the system parameters $\{\thetav,c,b_m,\forall m\}$ by  minimizing the MSE in \eqref{eq06}. To this end, CSI is required at both the receiver (\ie the PS, to optimize $c$ and $\thetav$)\footnote{According to \cite{wu2020intelligent}, the RIS phase shift vector $\thetav$ can be designed at the PS and sent to the RIS controller through a reliable backhaul link.} and the transmitters (\ie the devices, to optimize $\{b_m\}$). {In conventional wireless networks, CSI at the receiver (CSIR) can be efficiently estimated by uplink training in the existing solutions \cite{wang2020federated,liu2020reconfigurable}.} However, to obtain CSIT, the PS has to feed the CSI back to the devices through downlink channels. This not only incurs additional signaling overheads but also leads to imperfect model aggregation due to inevitable feedback errors \cite{MIMO_DTse}. 
To tackle the challenges in CSIT-based model aggregation, we propose a novel \emph{CSIT-free} model aggregation solution. Specifically, we assume perfect CSIR at the PS but no CSIT at the devices. 
In the proposed design, the devices transmit their signals all with full power (\ie $b_m=\sqrt{P_0} , \forall m$), and the RIS phase shift vector $\thetav$ is tuned to align the signals to achieve the desired sum $\sum_m p_m s_m[i]$. To facilitate our design, we first review the CSIT-based solution that minimizes the MSE in \eqref{eq06} in the following subsection.

\subsection{Preliminary: CSIT-Based Model Aggregation}\label{sec_3a}

Assume all the devices have perfect CSIT in this subsection. The PS and the devices can jointly optimize $\{\thetav,c,b_m,\forall m\}$ to minimize the MSE in \eqref{eq06}. 
Specifically, substituting \eqref{eq05} into \eqref{eq06}, we have
$\text{MSE}=\sum_{m=1}^M\Big| \frac{h_m(\thetav)b_m}{c} -p_m\Big|^2+{\sigma^2}/{|c|^2}\geq {\sigma^2}/{|c|^2}$,
where the last inequality is achieved if and only if ${h_m(\thetav)b_m}/c -p_m=0, \forall m$. Therefore, the optimal solution that minimizes the MSE must satisfy ${h_m( \thetav) b_m}/c =p_m,\forall m$. In other words, we must have
\begin{align}\label{eq08}
	b_m=\frac{p_m c}{ h_m(\thetav)}, \forall m.
\end{align}
By \eqref{eq08} and \eqref{eq03}, we have
$ |b_m|^2=\frac{|c|^2p_m^2}{|h_m(\thetav)|^2}\leq P_0, \forall m$, which is equivalent to
$|c|\leq \sqrt{P_0}\frac{|h_m(\thetav)|}{p_m}, \forall m$.
The minimum MSE under \eqref{eq08} is inversely proportional to $|c|^2$. Without loss of generality, we set $c$ as
\begin{align}\label{eq10}
	c= \sqrt{P_0}\min_{1\leq m\leq M}\frac{|h_m(\thetav)|}{p_m}.
\end{align}
Note that \eqref{eq10} minimizes the MSE and satisfies the power constraint.
Finally, the optimal RIS phase shift vector $ \thetav$ is the one that minimizes the MSE under \eqref{eq08} and \eqref{eq10} as
\begin{align}\label{eq11}
	\thetav=\argmin_{\thetav:\abs{\theta_l}^2=1,\forall l} 
	\frac{\sigma^2}{|c|^2}=\argmin_{\thetav:\abs{\theta_l}^2=1,\forall l}  \max_{1\leq m\leq M} \frac{p_m^2}{|h_m(\thetav)|^2}.
\end{align}
The optimization in \eqref{eq11} is generally non-convex. Existing solutions that approximately solve \eqref{eq11} can be found in \cite{wang2020federated,liu2020reconfigurable}. {We note that CSIT is critical in computing \eqref{eq08}. In the subsequent subsection, we propose a CSIT-free FEEL design based on the RIS.}

%

\subsection{Proposed CSIT-Free Model Aggregation}
From the previous subsection, we see that ${h_m(\thetav)b_m}/c =p_m$ is required in order to force the weight mismatch error $\sum_{m}| {h_m(\thetav)b_m}/{c} -p_m|^2$ to be zero. In this subsection, we assume no CSIT is available at the devices. Different from \eqref{eq08} that achieves ${h_m(\thetav)b_m}/c =p_m$ by optimizing $\{b_m\}$, we tune $\thetav$ to satisfy ${h_m(\thetav)b_m}/c =p_m$ and at the same time minimize $\sigma^2/|c|^2$.
Specifically, {we set all the transmit scalars to be a constant independent of the channel coefficients. To fully exploit the transmit power, the transmit scalars $\{b_m\}$ are chosen to achieve the largest power under \eqref{eq03}, \ie $b_m=\sqrt{P_0}, \forall m$.\footnote{Without loss of generality, the phases of the transmit scalars are set to zero.}}
Then, our target is to achieve $h_m(\thetav)\sqrt{P_0}/c\approx p_m$  by tuning $\thetav$. 
By doing this, we have
\begin{align}\label{eq12-0}
	\widehat r[i]=\sum_{m=1}^M \frac{h_m(\thetav)\sqrt{P_0}}{c} s_m[i]+\frac{n[i]}{c}\approx \sum_{m=1}^M  p_m s_m[i]+\frac{n[i]}{c}.
\end{align}
From \eqref{eq12-0}, we see that the estimate $\widehat r[i]$ approximates the aggregated sum with an additional noise term ${n[i]}/{c}$. In order to minimize the associated noise power, we simultaneously enforce $h_m(\thetav)\sqrt{P_0}/c\approx p_m$ and maximize $|c|^2$  by solving the following optimization problem:
\begin{subequations}\label{eq12}
	\begin{align}
		\max _{c\neq 0,\thetav\in \Complex^{L\times 1}} &\quad |c|^2\\
		\text{s.t. }& \abs{\theta_l}^2=1,1\leq l\leq L,\\
		& \sum_{m=1}^M \abs{h_m(\thetav)-{cp_m}/{\sqrt{P_0}}}^2\leq \e\label{eq12c}.
	\end{align}
\end{subequations}
In \eqref{eq12c}, we ensure  $h_m(\thetav)\sqrt{P_0}/c\approx p_m$ for $\forall m$ with the approximation error constrained by a small pre-determined scalar $\e$. 

{In the following, we propose an efficient algorithm to solve the non-convex problem \eqref{eq12}.}
The original problem in \eqref{eq12} is equivalent to 
\begin{subequations}\label{eq13}
	\begin{align}
		\min_{\widetilde \thetav\in\Complex^{(L+1)\times 1},\widetilde \thetav\neq{ \bf 0}} &\quad - \abs{\widetilde \theta_{L+1}}^2\\
		\text{s.t. }& \norm{\Gv\widetilde \thetav+\fv}_2^2 \leq \e,\abs{\widetilde \theta_l}^2=1,1\leq l\leq L,\label{eq13c}
	\end{align}
\end{subequations}
where  $\widetilde \thetav=[\thetav^T,c]^T\in\Complex^{(L+1)\times 1}$, 
$	\Gv=\begin{bmatrix}
	\gv_1^T,&-p_1/\sqrt{P_0}\\
	\cdots&\cdots\\
	\gv_M^T,&-p_M/\sqrt{P_0}\\
\end{bmatrix}$, and
$\fv=[	h_{DP,1},\cdots,	h_{DP,M}]^T$.
Furthermore, by introducing an auxiliary variable $\tau$, Problem \eqref{eq13} can be recast as 
\begin{subequations}\label{eq15}
	\begin{align}
		\min_{{\bf v} \in\Complex^{(L+2)\times 1},{\bf v}\neq{ \bf 0}} &\quad -|v_{L+1}|^2\\
		\text{s.t. }& \abs{v_l}^2=1,l=1,2,\cdots,L,L+2,\\
		&{\bf v}^H\Rv{\bf v} +\norm{\fv}_2^2\leq \e\label{eq15c},
	\end{align}
\end{subequations}
where
$\Rv=\begin{bmatrix}
	\Gv^H\Gv&\Gv^H\fv\\
	\fv^H\Gv&0
\end{bmatrix}$, and ${\bf v}=[\tau\widetilde \thetav,\tau]^T$.
{Next, we adopt matrix lifting to tackle the non-convex problem \eqref{eq15}.}
Define $\Vv\triangleq{\bf v}{\bf v}^H\in\Complex^{(L+2)\times(L+2)}$ that satisfies $\Vv\succeq \bf 0$ and $\rank(\Vv)=1$. The problem in \eqref{eq15} is equivalent to 
\begin{subequations}\label{eq17}
	\begin{align}
		\min_{\Vv\succeq\bf 0,\Vv\neq{ \bf 0}} &\quad -V_{L+1,L+1}\\
		\text{s.t. }& V_{l,l}=1,l=1,2,\cdots,L,L+2,\\
		&\rank(\Vv)=1,\label{eq17c}
		\tr(\Rv\Vv) +\norm{\fv}_2^2\leq \e,
	\end{align}
\end{subequations}
where $\tr(\cdot)$ is the trace operator.
Here, we adopt DC programming to approximately solve \eqref{eq17}. Note that,  for any $\Vv\in \Complex^{(L+2)\times (L+2)}$ such that $\Vv\succeq\bf 0$ and $\Vv\neq{ \bf 0}$, we have $\tr(\Vv)\geq\norm{\Vv}_2$, where $\norm{\Vv}_2$ is the spectral norm of $\Vv$. Moreover, according to  \cite[Proposition 3]{FL_1}, we have
$	\rank(\Vv)=1\Leftrightarrow \tr(\Vv)-\norm{\Vv}_2=0$.
Following \cite{FL_1}, we apply this property and move $\tr(\Vv)-\norm{\Vv}_2$ into the objective function as
\begin{subequations}\label{eq18}
	\begin{align}
		\min_{\Vv\succeq\bf 0,\Vv\neq{ \bf 0}} &\quad \rho(\tr(\Vv)-\norm{\Vv}_2)-V_{L+1,L+1}\label{eq18a}\\
		\text{s.t. }& V_{l,l}=1,l=1,2,\cdots,L,L+2,\\
		&\tr(\Rv\Vv) +\norm{\fv}_2^2\leq \e,
	\end{align}
\end{subequations}
where $\rho>0$ is the penalty parameter.
In \eqref{eq18}, we obtain a rank-one solution when the nonnegative penalty term $\rho(\tr(\Vv)-\norm{\Vv}_2)$ is enforced to zero.

Finally, to tackle the non-convex term $-\rho\norm{\Vv}_2$ in \eqref{eq18a}, we apply majorization-minimization to iteratively linearize $-\rho\norm{\Vv}_2$. That is, for iteration $i=1,2,..., I_{\text{max}}$, we construct a surrogate function to approximate $-\rho\norm{\Vv}_2$ based on the current solution $\Vv^{(i)}$ by noting that
\begin{align}
	-\rho\norm{\Vv}_2&\leq 	-\rho\norm{\Vv^{(i)}}_2+\tr(\Vv\cdot\partial_{\Vv^{(i)}} (-\rho\norm{\Vv}_2))\nonumber\\
	&=-\rho\tr(\Vv \uv^{(i)}(\uv^{(i)})^H)-\rho\norm{\Vv^{(i)}}_2,\label{eq19}
\end{align}
where $\partial_{\Vv^{(i)}} (-\rho\norm{\Vv}_2))$ is the subgradient of $-\rho\norm{\Vv}_2$ w.r.t. $\Vv$ at  $\Vv^{(i)}$; and $\uv^{(i)}$ is the principle eigenvector of $\Vv^{(i)}$. In \eqref{eq19}, we have used the fact that $\uv_1\uv_1^H\in\partial_{\Vv^{(i)}} \norm{\Vv}_2$
\cite[Proposition 4]{FL_1}. Using the right hand side of \eqref{eq19} to replace $	-\rho\norm{\Vv}_2$ in \eqref{eq18a}, we obtain the following convex problem:
\begin{subequations}\label{eq20}
	\begin{align}
		\Vv^{(i+1)}=\argmin_{\Vv\succeq\bf 0,\Vv\neq{ \bf 0}} &\quad \rho\tr\left(\Vv\left( \Iv_{L+2}-\uv^{(i)}(\uv^{(i)})^H\right) \right)-V_{L+1,L+1}\label{eq20a}\\
		\text{s.t. }& V_{l,l}=1,l=1,2,\cdots,L,L+2,\\
		&\tr(\Rv\Vv)+\norm{\fv}_2^2 \leq \e.
	\end{align}
\end{subequations}

For $i=1,\cdots,I_{\text{max}}$, the constructed convex problem \eqref{eq20} can be solved by standard convex optimization solvers.
After solving \eqref{eq20}, we retrieve the solution ${\bf v}$ in \eqref{eq18} by Cholesky decomposition $\Vv={\bf v}{\bf v}^H$.
Finally, the solution to \eqref{eq12} is given by $c=v_{L+1}/v_{L+2}$ and $\thetav=[{\bf v}]_{1:L}/v_{L+2}$, where $[{\bf v}]_{1:L}$ denotes
the vector of the first $L$ elements in ${\bf v}$. 
\begin{table}[t]
	\caption{Comparisons of CSIT-based and CSIT-free model aggregation methods}\label{table1}
	\centering
	\medskip
	\begin{tabular}{l|l|l}
		\toprule
		&CSIT-based&CSIT-free\\
		&& (Proposed)
		\\
		\hline
		Transmit scalar
		$b_m$&\eqref{eq08}&$\sqrt{P_0}$
		\\
		\hline
		Receive scalar 
		$c$&\eqref{eq10}&Solution to \eqref{eq20}
		\\
		\hline
		RIS phase shifts 
		$\thetav$&Solution to \eqref{eq11}&Solution to \eqref{eq20}
		\\
		\hline
		Requiring CSIT & Yes &No\\
		\bottomrule
	\end{tabular}
\end{table}

{Concerning the computation complexity, we adopt the second-order interior point method to solve each problem in \eqref{eq20}. The complexity of the proposed optimization algorithm is $O(IL^7)$, where $I\leq I_{\text{max}}$ is the number of iteration in which the update in  \eqref{eq20} converges. The proposed method has the same complexity order as the CSIT-based method in \cite{wang2020federated}. Compared with the CSIT-based method in \cite{liu2020reconfigurable} whose complexity is $O(M^4)$, our design has higher complexity since we adopt matrix lifting to solve the formulated problem.
}
We summarize the proposed method in Table \ref{table1}. 
{The proposed method does not require CSIT and thus avoids the additional feedback delay and errors. In other words, our method consumes more computation time in exchange for avoiding CSIT feedback overhead compared with the CSIT-based method in \cite{liu2020reconfigurable}.} 

\section{Numerical Results}\label{sec_simu}
In this section, we conduct simulations to examine the performance of the proposed method.\footnote{More simulation results can be found in the extended version of this work (available at https://arxiv.org/abs/2102.10749).} We consider a system with $M=40$ devices and $L=110$ RIS elements.
We simulate an image classification task over the Fashion-MNIST dataset \cite{Xiao2017}. The learning model is a convolutional neural network comprising two $5\times 5$ convolutional layers with  $\text{stride}=2$ (each followed with $2\times2$ max pooling), a batch normalization layer, a fully connected layer with $50$ neurons and ReLu activation, and a softmax output layer (total number of parameters $d=92,208$). The loss function $F(\cdot)$ is the cross-entropy loss. 
Each device has $1500$ training samples
(\ie $p_m=1/40$ for $\forall m$). 
We set $P_0=-20$ dB and $\sigma^2=-120$ dB. The PS and the RIS are located at $(-50,0,10)$ and $(0,0,10)$, respectively. The locations of the devices are i.i.d. drawn in $\{(x,y,0):-20\leq x\leq 0,0\leq y\leq 10\}$. We adopt a Raleigh fading channel model.
The direct channel path loss is given by $G_{\text{PS}}G_{\text{Device}}(\lambda_c/(4\pi d))^3$, where $d$ is the link distance, $G_{\text{PS}}=4.11$ dBi (or $G_{\text{Device}}=0$ dBi) is the PS (or device) antenna gain; $\lambda_c=(3*10^8m/s)/(815 \text{MHz})$ is the carrier wavelength. The RIS path loss is given by \cite[eq. (8)]{tang2019wireless}, where the RIS antenna gain is $4.11$ dBi; the size of each RIS element is $0.1\lambda_c\times 0.1 \lambda_c$; and the reflection amplitude $A$ is $1$. For FedAvg, we set $B=100$,  $\eta=10^{-4}$, $E=5$ for the case with i.i.d. data and $E=1$ for the case with non-i.i.d. data. For the algorithm in \eqref{eq20}, we set $\e=10^{-2}$, $\rho=10$, and $I_{\text{max}}=100$.
\begin{figure}[!t]
	\centering
	\includegraphics[width=5in]{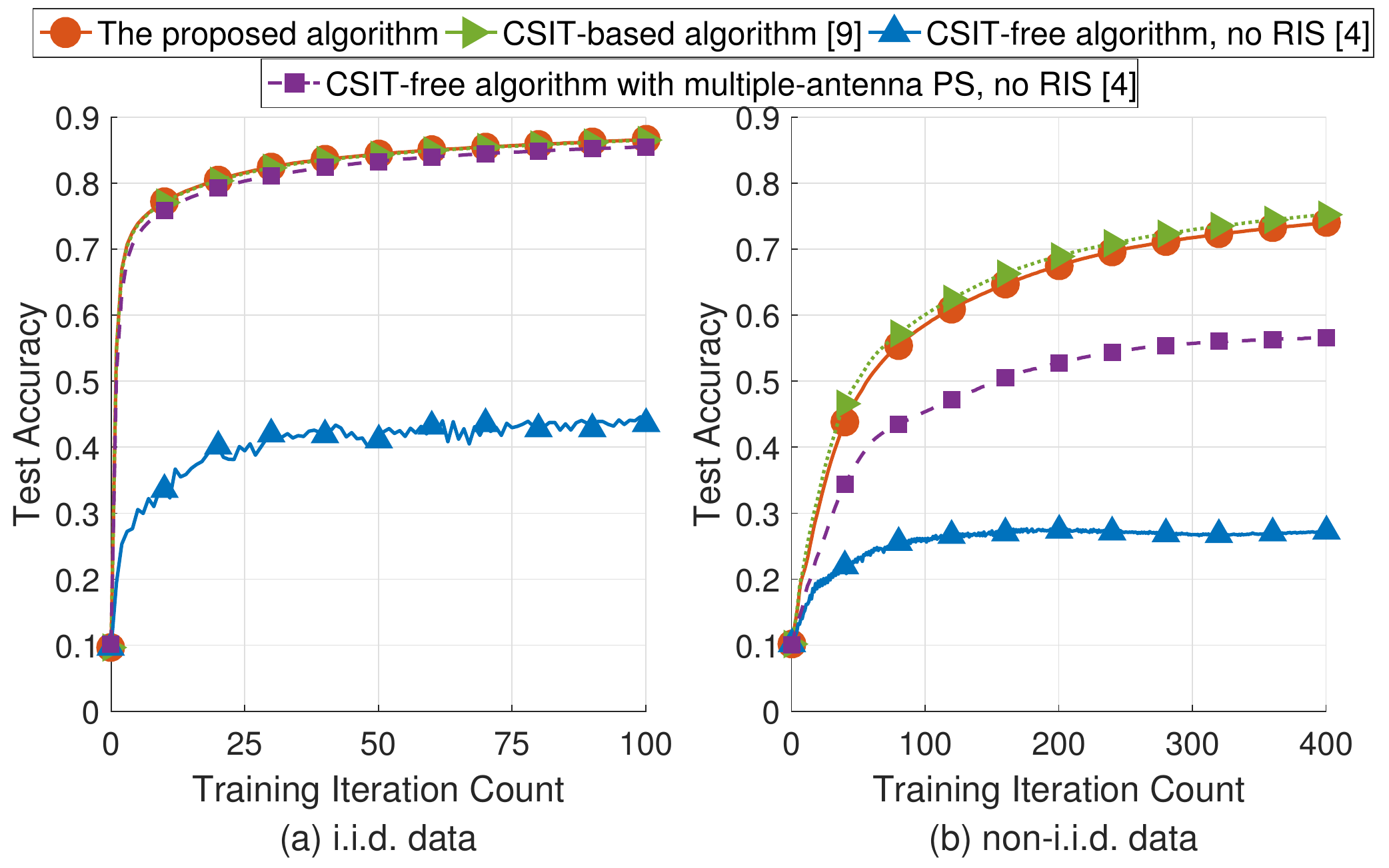}
	\caption{Test accuracy versus training iteration under (a) i.i.d. and (b) non-i.i.d. data distributions with $M=40$ and $L=110$. 
		%
	}
	\label{Fig1}
\end{figure}
\begin{figure}[!t]
	\centering
	\includegraphics[width=5in]{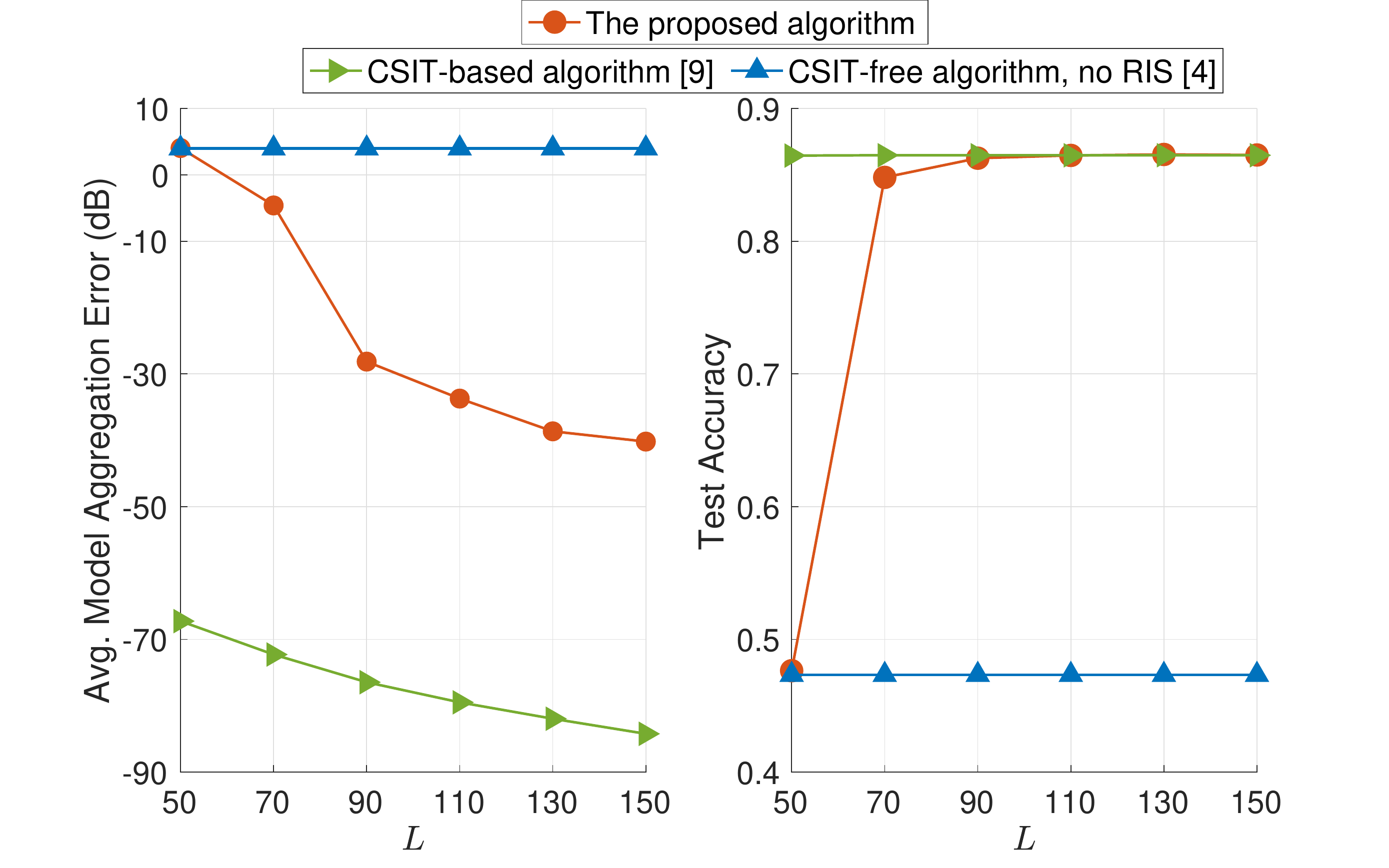}
	\caption{Aggregation MSE (left) and test accuracy (right) versus $L$ with i.i.d. data, $M=40$, and $T=100$.}
	\label{Fig2}
\end{figure}
In Fig. \ref{Fig1}, we plot the test accuracies of the proposed method and the CSIT-based method  \cite[Algorithm 1]{liu2020reconfigurable}. {The RIS has $L=110$ elements and the PS has one antenna}.
{We consider two data distributions: (a) i.i.d. distribution where local samples are i.i.d. drawn; and (b) non-i.i.d. distribution where $60,000$ samples are sorted by label and divided into $160$ subsets of size $375$ and each device is assigned $4$ subsets.}
%
The CSIT-free design without RIS from \cite{amiri2020blind} is also included for comparisons.  {The blue and purple curves represent the performance of the method in \cite{amiri2020blind} with one receive antenna and $110$ receive antennas at the PS, respectively.}
We see that the proposed algorithm achieves a similar accuracy to the CSIT-based one. 
In contrast, when the PS only has one receive antenna, the method in \cite{amiri2020blind} suffers from large inter-user interference. {Comparing the proposed algorithm (the orange curve) with the multiple-antenna-based method (the purple curve), we find that RIS \emph{passive} beamforming achieves a better accuracy than active beamforming via receive antennas. This result verifies the efficiency of the proposed RIS-based CSIT-free algorithm as a passive RIS element costs much less power than an active receive antenna \cite{wu2020intelligent}.  Even with non-i.i.d. training data, our method still achieves a similar accuracy with the CSIT-based baseline  \cite{liu2020reconfigurable}.
}

Fig. \ref{Fig2} plots the model aggregation MSE and the test accuracy under various RIS sizes $L$ with  i.i.d. data distribution. We consider a single-antenna PS for all the algorithms. The model aggregation  error in the left subfigure is defined as the MSE between the global model change $\sum_m p_m \Delta\wv_{m,t}$ and its estimate.
On one hand, compared with the CSIT-based method, the proposed method has a relatively larger MSE (\eg when $L=90$, the MSEs of our method and the CSIT-based one are $-28$ dB and $-76$ dB, respectively). However, the difference between these two methods in terms of test accuracy becomes indistinguishable when $L\geq 90$. This is because the proposed method has already achieved a relatively small aggregation error, which has a negligible impact on the learning performance. On the other hand, when $L< 90$, the considered optimization problem \eqref{eq12} becomes infeasible, and the proposed solution fails to align the channels. We conclude from Fig. \ref{Fig2}  that our proposed CSIT-free method performs as well as the CSIT-based one in terms of test accuracy provided that the RIS is sufficiently large.  


\begin{figure}[!t]	
	\centering
	\includegraphics[width=4 in]{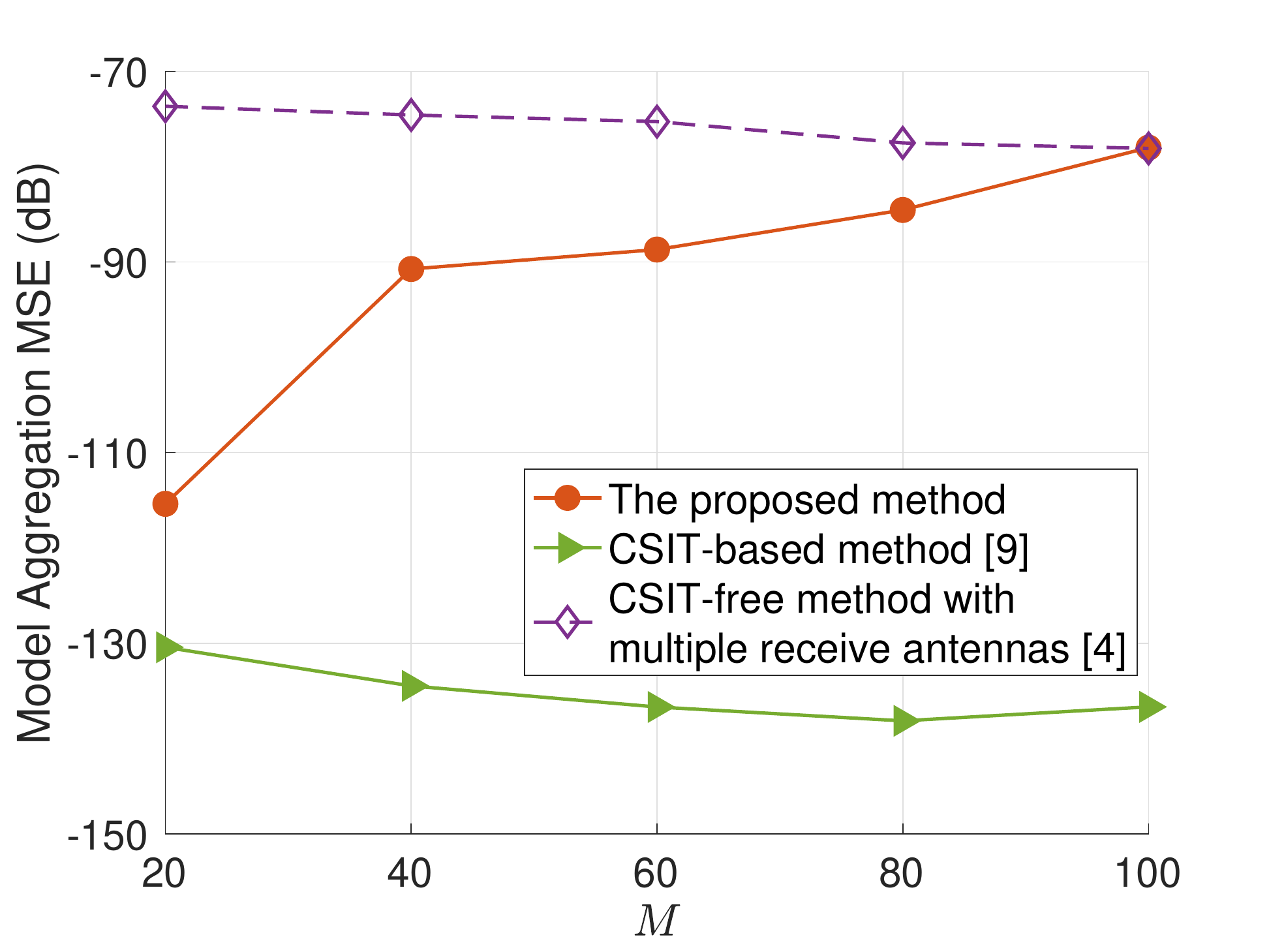}
	\caption{Model aggregation MSE versus $M$ with $L=150$. The orange and green curves use a single-antenna PS with $110$ RIS elements, and the purple curve uses  $110$ receive antennas with no RIS.}
	\label{M}
\end{figure}

Next, we study the performance versus the number of devices $M$. We set $L=150$ and $p_m=1/M$. Each device has $\lceil \frac{60000}{M}\rceil$ training samples, where $\lceil \cdot\rceil$ is the ceiling function. The other setting parameters are the same as those in Fig. 2(a) of the manuscript. Fig. \ref{M} plots the model aggregation MSE under different values of $M$. Our method (the orange curve) and the CSIT-based baseline from [9] (the green curve) adopt a single-antenna PS and $110$ RIS elements. For a fair comparison, the method from [4] (the purple curve) adopts $110$ receive antennas with no RIS. We find that
\begin{itemize}
	\item The MSEs of the baselines in [4], [9] are robust against $M$. More specifically, we observe that their MSEs slightly decrease as $M$ increases. This is because the MSE of over-the-air computation is roughly dominated by the device with the worst channel condition. Although adding more devices potentially brings in devices with weak channel conditions, the baselines in [4], [9] effectively mitigate their communication errors and maintain a relatively stable estimation. Meanwhile, as $M$ increases, the weight of each device $p_m=1/M$ decreases, and hence the corresponding weight of the weak devices  also decreases in model aggregation. The coupling  of the above two effects leads to a slightly better MSE. 
	\item The MSE of our method increases as $M$ increases. This is because our method relies only on RIS phase shifts $\thetav$ to align the channels to be proportional to the weights $\{p_m\}$, \ie 
	\begin{align}\label{eq0222}
		h_m(\thetav)\propto p_m=1/M,
	\end{align}
	where $h_m(\thetav)$ is the cascaded channel of device $m$ (as a function of $\thetav$).  Note that the passive RIS only tunes the phase shifts but cannot amplify the impinging signals. When more devices are considered, there is a larger probability that a device with a weak channel condition exists. In this case, our solution needs to correspondingly lower down  the channel gain of other devices  in order to achieve \eqref{eq0222}. Therefore, the resulting MSE inevitably increases.
	\item Although the aggregation MSE of our method decrease with $M$, our simulation shows that, when $20\leq M\leq 80$, all the methods, including ours, achieves good test accuracies ($\approx 0.86$), and there is no noticeable gap between our method with the baselines in terms of test accuracy. This is because all the methods achieve sufficiently small MSEs ($<-70$ dB), which have no harm to test accuracy. When $M=100$, our method has a small test accuracy degradation ($\approx 0.855$).
	When $M$ is extremely large, we need to increase the number of RIS elements $L$ to ensure that our method still works well; see Fig. 3 in the manuscript.
\end{itemize}
We conclude from Fig. \ref{M} that our method works well with a small or medium number of devices. In contrast, the existing methods based on CSIT or MIMO techniques are more robust against $M$. However, Fig. 3 shows that \emph{increasing the number of RIS elements is an efficient way to improve the communication performance of our method as RIS elements consume very low power}.
\begin{figure}[!t]	
	\centering
	\includegraphics[width=4 in]{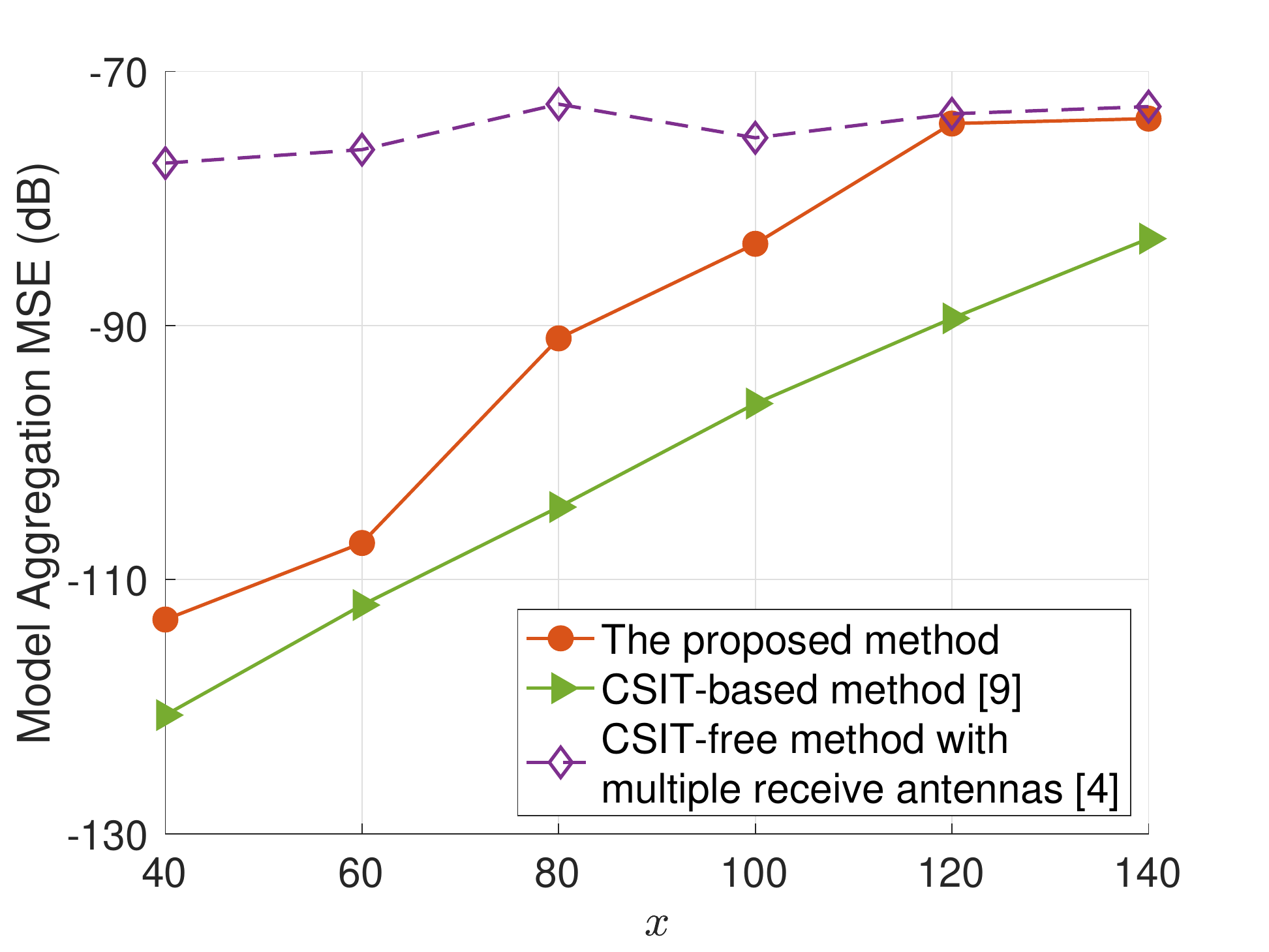}
	\caption{Model aggregation MSE versus $x$ with $L=150$ and $M=40$. The orange and green curves use a single-antenna PS with $110$ RIS elements, and the purple curve uses  $110$ receive antennas with no RIS.}
	\label{distance}
\end{figure}

Next, we study the effect of the communication distance. We set $L=150$ and $M=40$. The PS and the RIS are located at $(-50,0,10)$ and $(0,0,10)$, respectively. The devices are uniformly located at $[x-20,x]\times [0,10]$.  The path loss models are given in Section IV of the manuscript. Generally, a larger value of $x$ means a longer communication distance, hence a smaller path loss coefficient. Fig. \ref{distance} plots the MSE versus the value of $x$. When $x$ is small, our method achieves a smaller MSE than the baseline in [4], verifying the effectiveness of our proposed RIS-assisted design. 
When $x$ is large, the MSE of our method approaches that of [4]. This shows that the effect of passive RIS phase shifts becomes limited when the channel gain decreases because a passive RIS cannot amplify the signal power. We also note that increasing the number of RIS elements can efficiently combat the effect of deep path attenuation; see Fig. 3 in the manuscript.

\begin{figure}[!t]
	\centering
	\includegraphics[width=4in]{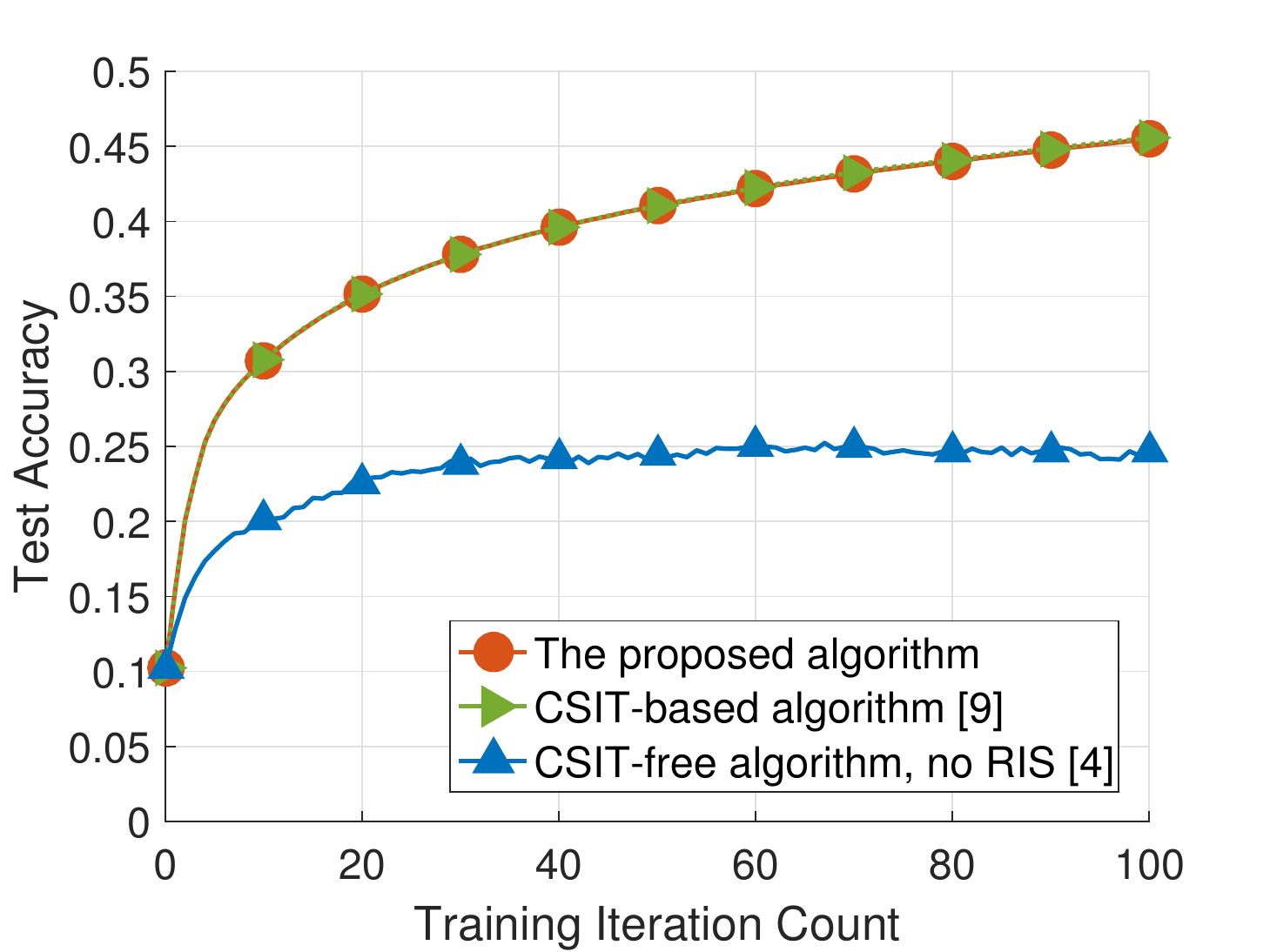}
	\caption{Test accuracy versus training iteration for CIFAR-10 image classification with i.i.d. data distribution, $M=40$ and $L=110$.}
	\label{g}
\end{figure}
Then, we examine our method for another popular dataset, namely CIFAR-10 \cite{cifar10}. We use the convolutional neural network setup in \cite{FEDSGD} to train the classification model. Fig. \ref{g} plots the learning accuracy under this dataset where we consider an i.i.d. data distribution as in Fig. \ref{Fig2}. We find a similar result  with that in the Fashion-MNIST dataset: The proposed algorithm achieves performs as well as the CSIT-based method.

\begin{figure}[!t]
	\centering
	\includegraphics[width=4in]{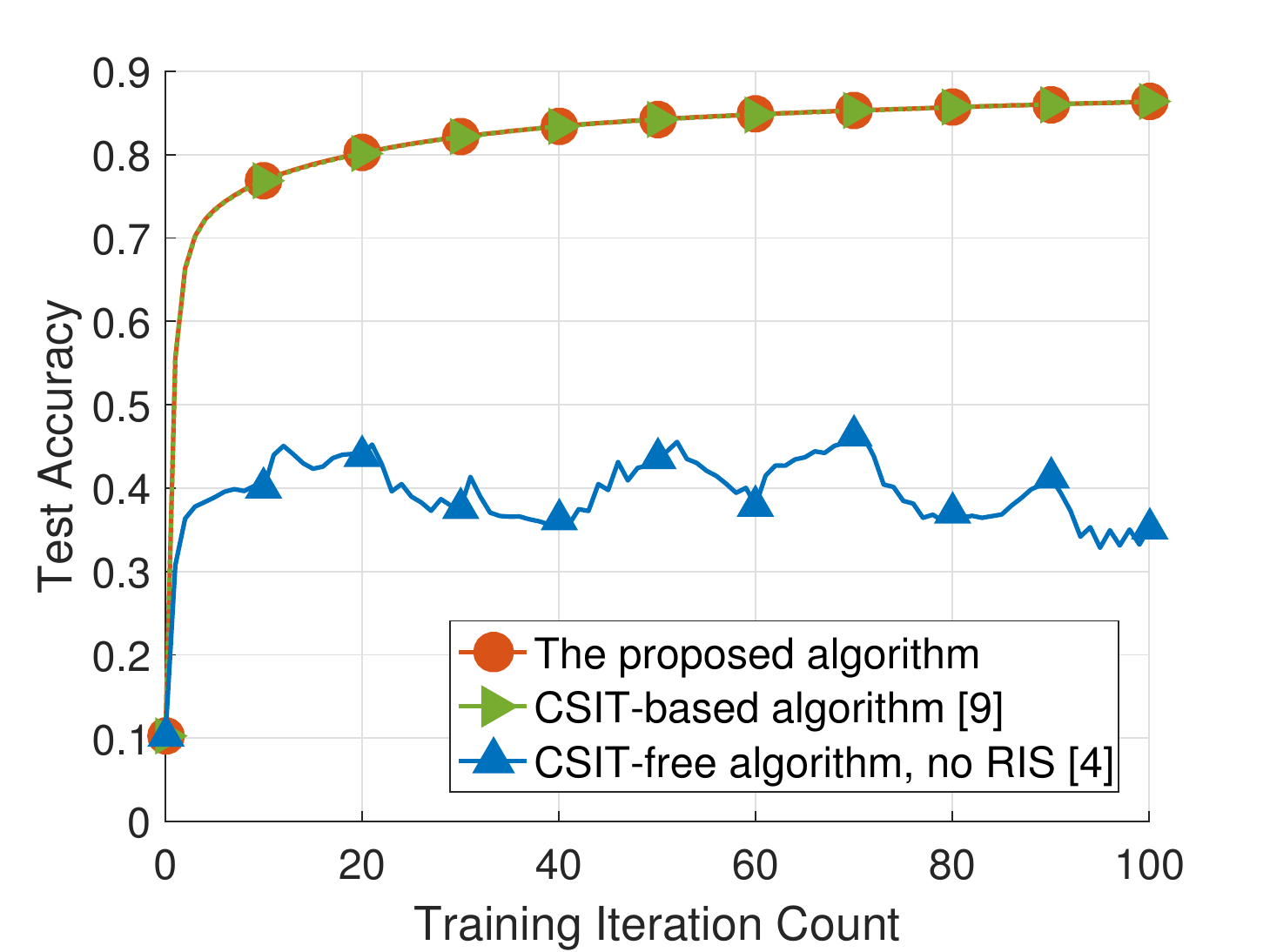}
	\caption{Test accuracy versus training iteration with time-varying channels, i.i.d. data, $M=40$, and $L=110$.}
	\label{j}
\end{figure}

Moreover, we study another simulation setup with time-varying channels to demonstrate the effectiveness of our method in this scenario. 
Specifically, we assume that the small-scale fading channel coefficients of all the channels vary independently every 10 training rounds. Therefore, with $T = 100$ training rounds, the small-scale
fading coefficients change $10$ times, and we need to update our RIS design everytime when the channel changes. We plot the learning accuracy of the proposed algorithm with time-varying channels in Fig. \ref{j}. Except for the small-scale fading coefficients, the other parameters are exactly the same as in Fig. 2(a). We find that our method still achieves an accuracy 
close
to the CSIT-based baseline, demonstrating the robustness of our algorithm under time-varying channels. 

\begin{figure}[!t]	
	\centering
	\includegraphics[width=4 in]{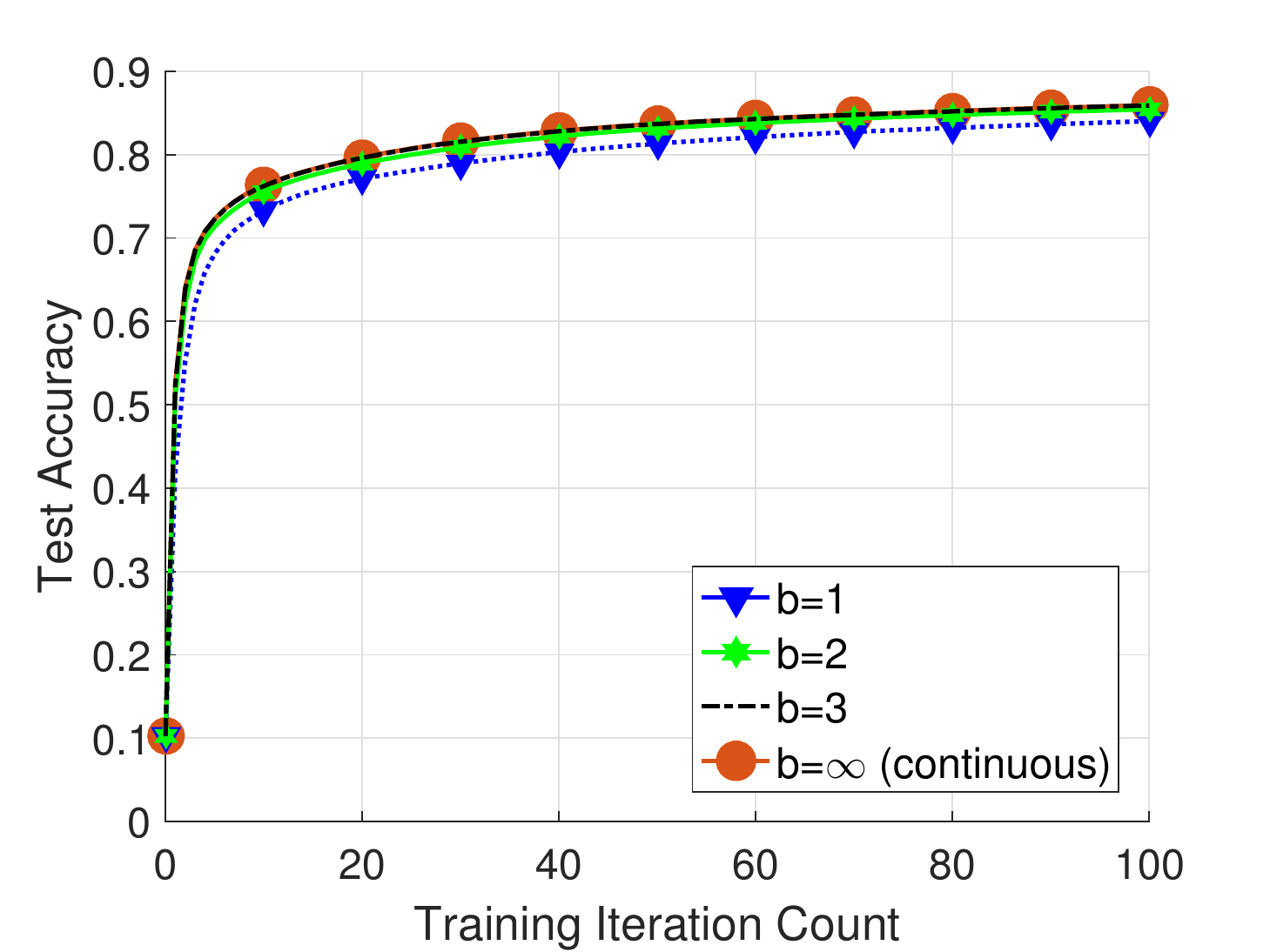}
	\caption{FL accuracy of the proposed algorithm with different discrete phase shift levels and i.i.d. data. We set $T=100$, $M=40$, and $L=110$.}
	\label{dis2}
\end{figure}

We consider a discrete-phase-shift RIS model for $\thetav$ with low-bit reflecting elements. Specifically, we assume that each phase shift $\theta_l, l=1,2,\cdots,L$ can take only a finite number of discrete values. The available RIS phase shift vector is given by
\begin{align}\label{eq072}
	\theta_l\in\Theta_b\triangleq \{\theta=e^{j\frac{2\pi i}{2^b}}:0\leq i\leq 2^b-1\},\forall l,
\end{align}
where $b$ is the number of phase shift bits. Note that $\Theta_b$ in \eqref{eq072} converges to the continuous feasible set with $b\to \infty$.
We here accommodate our original solution to the case with discrete RIS phase shifts by projecting the solution to \eqref{eq072}. 
First, we treat $\thetav$ as if $b=\infty$, \ie 
$\thetav$ can take continuous values. Then, we use the proposed algorithm in eq.(18) to optimize $\thetav$. Finally, we project the solution to the feasible set given in \eqref{eq072}. 
Fig. \ref{dis2} plots the learning accuracy of our method under various choices of $b\in\{1,2,3,\infty\}$. Except for the model on $\thetav$, other system parameters are the same as in Fig. 2(a). 
We find that our method with $b=2$ or $3$ yields the same performance with the continuous case ($b=\infty$). This is because our proposed RIS-based design has already achieved a very small aggregation error, and hence the performance loss due to the discrete projection has an imperceptible influence on the final learning accuracy. 
When $b=1$, the loss from the low-bit projection becomes relatively large, and we observe a $0.01$ accuracy degradation in this case. We conclude from Fig. \ref{dis2} that our algorithm works well with $b\geq 2$ and the learning accuracy is not sensitive to the value of $b$.

Finally, we numerically explore our proposed RIS-assisted model  aggregation method with downlink transmission errors. Specifically, 
\begin{figure}[!t]	
	\centering
	\includegraphics[width=4 in]{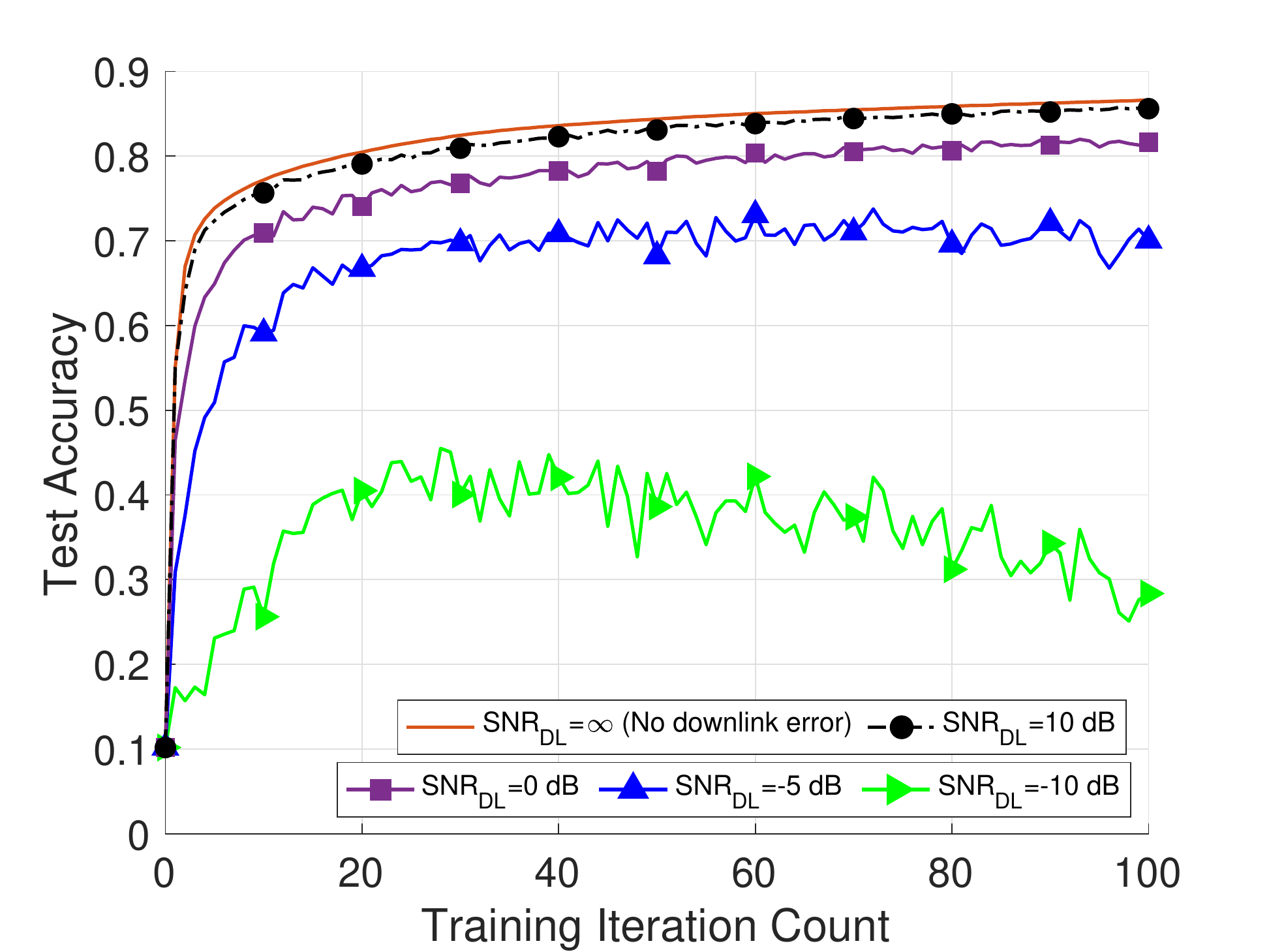}
	\caption{Test accuracy of the proposed method under different levels of downlink error. A higher value of  $\text{SNR}_{\text{DL}}$ means less downlink error.}
	\label{erroracc}
\end{figure}
in the model broadcasting procedure of the $t$-th learning round, we assume that the received global model at each device is the accurate global model $\wv_t$ plus some additive Gaussian noise. That is, the received global model at each edge device is given by
\begin{align}\label{eq033}
	\widetilde \wv_t=\wv_t+\nv_{\text{DL},t},
\end{align}  
where the entries of $\nv_{\text{DL},t}$ are drawn from $\CN(0,\sigma^2_{\text{DL}})$.
This noisy global model $\widetilde \wv_t$ is used in local model training, and the resulting local model change is uploaded to the server for model aggregation as discussed in the manuscript. The downlink SNR to \eqref{eq033} is defined as $\text{SNR}_{\text{DL}}\triangleq 1/\sigma^2_{\text{DL}}$. 
We plot the test accuracies of the proposed method with various levels of $\text{SNR}_{\text{DL}}$ in Fig. \ref{erroracc}. We find that our method performs reasonably well when $\text{SNR}_{\text{DL}}\geq 0$ {dB}. When $\text{SNR}_{\text{DL}}< 0$ {dB}, the downlink noise $\nv_{DL,t}$ overwhelms the model vector $\wv_t$, and the learning accuracy becomes diverge.
Note that RISs have been adopted to enhance the broadcasting quality in conventional communication systems. We envision here that RIS phase shifts can also be introduced to improve the downlink SNR to avoid the accuracy degradation. We will study this direction in our future work.



\section{Conclusions}
In this letter, we studied CSIT-free model aggregation for RIS-assisted FEEL. We adopted the RIS phase shifts to align the cascaded channels with the aggregation weights.  
We developed a novel algorithm to solve the resulting optimization problem. Finally, simulations on image classification show that, despite the lack of CSIT, our algorithm achieves a similar accuracy as the existing CSIT-based solution when the number of RIS elements is sufficiently large. 
Promising extensions to this work include joint passive and active beamforming, unified uplink and downlink model transmissions with RISs, and robust design against CSI errors. 
{In particular, except for uplink model aggregation, model broadcasting through downlink channels also suffers from performance loss due to the existence of communication noise \cite{nosiydl}. RISs can be adopted to enhance downlink model transmissions, which requires new designs on RIS phase shifts.
}

\ifCLASSOPTIONcaptionsoff
\newpage
\fi
\ifhavebib
{
	\bibliographystyle{IEEEtran}
	\bibliography{ref}
}
\else{
}
\fi
\end{document}